\documentclass[preprint2]{aastex}
\usepackage{psfig}
\slugcomment{Submitted to the Astrophysical Journal}

\shorttitle{Luminous compact galaxies}
\shortauthors{Hammer et al.}
\begin{document}
\title{Luminous compact galaxies at intermediate redshifts: progenitors of
bulges of massive spirals ?}
\author{Fran\c cois Hammer\altaffilmark{1} and Nicolas Gruel}
\affil{DAEC, Observatoire de Paris-Meudon, 92195 Meudon, France}
\author{Trinh X. Thuan}
\affil{Department of Astronomy, University of Virginia, 
Charlottesville, VA 22903}
\author{Hector Flores}
\affil{Service d'Astrophysique, CEA-Saclay, 91191 Gif-sur-Yvette, France}
\and
\author{Leopoldo Infante}
\affil{Universidad Catolica, Santiago, Chile}


\altaffiltext{1}{Visiting Astronomer, Very Large Telescope,
 operated by The European Southern Observatory}

\begin{abstract}
VLT spectra of 14 luminous compact galaxies (LCGs) reveal 
strong metallic absorption line sytems as well as narrow and intense emission
lines. Their gas extinction is found to be large 
(A$_V$ $\sim$ 1.5 mag) leading
to an upward revision of their star formation rate (SFR) to  
an average value of $\sim$ 40 $M_{\odot} yr^{-1}$. Large extinction values are
also supported by the large rate of detection in one field observed by ISO.
Gas metal abundances in LCGs have about half the solar value. 
LCG absorption spectra can be synthesized with a mix of a few 
Gyr old and relatively metal-rich (generally solar to over-solar values) 
stellar population and a younger stellar population 
($<$ 5$\times 10^{8}$ years) having a   
metal abundance similar to that of the gas. \\

We argue that LCGs are the progenitors of present-day spiral bulges.
LCGs have masses and light concentrations similar to those of present-day
bulges. They could have been formed entirely during a 
period of a few
Gyr prior to the epoch of their observations 
if the star formation has been sustained at the 
observed rate. As in present-day galactic bulges, LCG 
stars show a wide range of abundances.
Thus, observing LCGs allows us to directly witness an important 
stage in the formation of a massive galaxy, the building of the bulge 
prior to that of the disk. The gas needed to feed
the observed star formation is likely to be falling in from
the outskirts of the galaxy, being tidally pulled out from 
interacting companion galaxies. An infall scenario naturally explains the
 gas metal abundance which is generally lower than that of the older stellar
 component. At least for the strongest star-forming
LCGs, there is clear imaging evidence for the presence of companions.
 Some LCGs also show evidence for the beginning of a disk formation.
\\
If the above scenario holds for most LCGs, we estimate that 
at least 20\% of present-day spiral galaxies
have formed the bulk of their stars at relatively recent epochs, during the
last 8-9 Gyr, at redshifts less than $\sim$ 1. 
Since they are heavily extincted, we predict their
 IR luminosities to be relatively large, 
around $L_{IR} $=$ 10^{11}$ $L_{\odot}$, i.e. near or
slightly below the luminosities 
of the galaxies detected by ISO in the same redshift range.
Taking into account the integrated IR luminosity of the LCG 
galaxy population can lead to a significant upward revision of the
cosmic SFR density in the redshift range from 0.5 to 1.
\end{abstract}



\newpage

\section{Introduction}
The evolution of UV and IR luminosity densities can be understood as a
decrease by a factor of $\sim$ 6 of the global star formation density
from the epoch corresponding to redshift z=1 to the present epoch
(Lilly et al. 1996; Flores et al. 1999). These luminosity densities
evolve similarly and contribute equally to the star formation in that
redshift range, if one assumes that the shapes of the UV and IR
luminosity functions to be similar to the local ones (Hammer
1999). Integration of the derived global star formation implies that
between half and two thirds of the present-day stars have been formed
since z=1. This is somewhat in contradiction with the primordial
collapse scenario in which the most massive galaxies were formed at
much earlier epochs, at redshifts larger than $\sim$ 3 (e.g. Renzini
and Cimatti 1999).  The latter scenario is supported by the apparent
non-evolution in the number density of large disks (Lilly et
al. 1998), and by its success in reproducing galactic disk evolution
with chemical evolution models normalized to local galaxies (Boissier
and Prantzos 2000).
\\
 The study of the luminosity density evolution relies on observations
 of luminous galaxies with $M_{B} <$ -20 in the optical, and with
 $L_{bol} >$ 2 $\times$ $10^{11}$ $L_{\odot}$ in the IR.  The two main
 galaxy populations responsible for the observed decline in the star
 formation density are, in decreasing importance, the luminous IR
 galaxies (Flores et al. 1999) and the luminous compact galaxies (LCG)
 (Guzman et al. 1997; Lilly et al. 1998).  The former are generally
 large and massive galaxies found mostly in interacting systems
 (Hammer, 1999), and which show star formation rates (SFR) exceeding
 50 $M_{\odot}yr^{-1}$. They represent $\sim$ 4\% of the field galaxy
 population.  The latter are much more numerous and contribute to more
 than 40\% of the UV luminosity density at z$\sim$ 1 (Guzman et
 al. 1997; Hammer \& Flores 1998). They constitute the most rapidly
 evolving galaxy population seen in the UV.  The properties of these
 LCGs are still largely unknown. They are sometimes so compact that
 their nucleus is barely resolved in the WFPC2 HST images, which
 prevents their classification on the basis of their luminosity
 profile.\\
\\

Koo \& Kron (1981) first drew attention to these compact galaxies
during their search for faint QSOs on the basis of colors and
compactness.  They identified a sparse population (the surface density
is $\sim$ 30 sources per square degree) of very blue compact galaxies
with z $\le$ 0.7 and exhibiting narrow emission lines.  Using the Keck
HIRES spectrograph, Koo et al. (1995) demonstrated that the emission
lines of LCGs have a Gaussian profile, and suggested that they may be
progenitors of local spheroidal galaxies through fading by $\sim$ 4 to
5 magnitudes. More recently, Guzman et al. (1997) and Phillips et
al. (1997) have identified a large population of emission-line compact
galaxies in the flanking fields around the Hubble Deep Field (HDF,
Williams et al.  1996) at z $\le$ 1, comprising $\sim$ 20\% of the
field galaxy population.  They found the [OII]$\lambda$3727
emission-line velocity widths of the compact galaxies to range between
35 and 150 km s$^{-1}$ (Phillips et al. 1997), comparable to those of
local dwarf galaxies, although the compact galaxies are 10 to 100
hundred times more luminous (up to a few $L^{\star}$). The
similarities between the HDF compact galaxies and those from Koo and
Kron (1981) have led Guzman (1999) to suggest that blue compact
galaxies at intermediate redshifts may be the progenitors of local
spheroidal or irregular galaxies through fading by $\le$ 4 magnitudes.
If true, according to Guzman (1999), these should have extremely low
M/L ratio, and this scenario would imply a severe downwards revision
of the amount of star formation at z$\sim$ 1, since the baryonic
content in irregulars (or spheroidals) is very small (Fukugita et
al. 1998).\\
\\

We develop in this paper a different scenario for the nature of the
brightest LCGs, which are major contributors to the observed decline
of star formation density since z=1.  In our picture, LCGs are not the
progenitors of present-day spheroidal and irregular galaxies, but
rather of today's spiral bulges which built up gradually over time by
a series of mergers of smaller units.  There have been several
previous observations that have hinted at that scenario, but their
meaning was not realized at the time. Thus, Phillips and al. (1997)
found that a large fraction of galaxies selected by their compactness
appeared to be small but bright spirals.  Even adding a color
criterion is not sufficient to distinguish between old and young
stellar populations: Schade et al. (1999) found that fully a third of
the distant ellipticals they studied show star formation and colors as
blue as those of blue compact galaxies.  The merging scenario comes
naturally as the merging rate was about ten times higher at z=1 than
it is today (Le F\`evre et al. 2000). Moreover, Guzman (1999) and
Hammer (1999) have noticed that a large fraction of compact galaxies
selected from the HDF and Canada-France redshift survey (CFRS) fields
are in interacting systems, with the presence of complex and distorted
features surrounding their cores.  Finally, the LCGs in the CFRS
fields have near-IR and visible colors very similar to those of other
luminous non-dwarf galaxies, with values apparently not consistent
with those produced by a strong and short burst of star formation in a
low-mass object (Hammer et al. 1997).  We present next new
observational data which support and sharpen the above scenario.\\ In
the following, $h_{50}$=$H_{0}$/50 and if not specified we adopt $h_{50}$=1 
and $q_{0}$=0.5.

\section{The sample: selection criteria}
Our LCG sample was selected from two galaxy fields studied in the
CFRS: the CFRS 0300+00 (Hammer et al. 1995) and the CFRS 2230+00
(Lilly et al. 1995) fields.  To define our sample, we apply 3
selection criteria. First is the compactness criterion, as determined
from deep HST images (see Brinchmann et al. 1998), or from deep
ground-based images obtained by the CFRS team, with an image quality
better than 0".8 FWHM and a pixel size of 0.207". It is based on
either a small size: 

$r_{1/2}$ $\le$ 5 $h_{50}^{-1}$ kpc, where $r_{1/2}$ is the half-light radius 
and $h_{50}$ the Hubble constant in units of 50 km s$^{-1}$ Mpc$^{-1}$

or on light concentration as defined by the parameter:
 
$\delta$ I = $m_{I}(15h_{50}^{-1}kpc)$ - $m_{I}(5h_{50}^{-1}kpc)$ $\le$ 0.75 mag (Hammer and Flores 1998).\\
\\

Imagery from HST has been analysed with SExtractor software package
(Bertin and Arnouts, 1996). The detection threshold was set to3.5
$\sigma$ and a 35 pixel annulus was used for sampling the background.
The minimum contrast parameter for deblending was set to 0.005, and
for each source with an identified companion at more than 2" (18 kpc
at z=0.5), an image with a mask superimposed on the companion has been
generated. The background level at the mask location has been set
using the background mean level, and the mask size was set from the
derived Kron radius of the companion (Bertin and Arnouts 1996).  This
procedure has the advantage of eliminating systematic biases against
compact sources with companions beyond 2". For HST images, 15
apertures with diameters varying from 0".2 to 3".0 with increasing
steps of 0".2, were used to derive aperture photometry. The half-light
radius has been derived assuming that the isophotal magnitude
generated by SExtractor was a good approximation to the total
magnitude of the source.  Calculation of the light concentration
parameter was done after interpolating to the luminosities enclosed
within the 5 kpc and 15 kpc radii.  Ground-based images were analyzed
in a similar way, but a 20-pixel annulus was used for background
sampling. Another problem with images obtained from the ground come
from background fluctuations.  For the majority of the objects,
aperture magnitudes are increasing smoothly with aperture
radius. Those that show a magnitude decrease larger than the
measurement errors have been excluded from the sample.\\
\\

The above selection criteria are equivalent for a face-on disk or a
spheroidal component with a r$^{1/4}$ profile since for a $r_{1/2}$
$\le$ 5 kpc source, $m_{I}(15h_{50}^{-1}kpc)$ is a good approximation of the
total source luminosity.  Figure 1 demonstrates this equivalence for
sources imaged with the HST.  The bottom panel shows that the light
concentrations calculated from HST and CFHT images for the same object
correlate well, the correlation coefficient being 0.73. We note
however that the average value of $\delta$ I from ground images is
larger by 0.1 magnitude as compared to that from the HST images. This
can be easily explained by the broadening of the profile due to seeing
conditions, and we have applied a slight empirical correction to the
values derived from the ground.  The correlation between ground and
HST data is found to be better for the concentration parameter (upper
panel) than for the half-light radius. We interpret this to be related
to the uncertainties in the calculation of the isophotal magnitudes
for ground-based images, because background fluctuations affect them
more than aperture magnitudes. As a result, we have decided to adopt
the concentration parameter as the selection criterion for compact
galaxies.  Examination of images of the selected compact galaxies
shows that the compactness selection criterion has also picked out a
few highly inclined disks with a large major axis extent. These were
removed from the sample using the axis ratio calculated by SExtractor.
One can also notice from Figure 1 that $\sim$ 20\% of the sources are
found to be compact on the basis of their HST photometry, although
they appear to be extended from the ground. Examination of their
images show that they are mostly sources with nearby companions
(within 3") for which the deblending procedure applied to ground-based
images has failed. In other words, our procedure applied to
ground-based images appears to be more restrictive than when applied
to higher spatial resolution HST images.  A selection from ground
images will inevitably generate a sample of galaxies slightly biased
against galaxies with nearby companions.\\
\\

Second, we apply a luminosity and redshift criterion. Only galaxies
with $M_{AB}$(B) $\le$ -20, and in the redshift range from 0.45 to 0.8
are included.  The luminosity cut-off is motivated by the fact that
the reported evolution of the rest-frame UV luminosity density is
based on observations of galaxies with $M_{AB}$(B) $\le$ -20 (Lilly et
al. 1996), while the redshift range is the one in which the CFRS is
considered complete for luminous galaxies (see Hammer et al. 1997).\\

Third, we include only LCGs with known [OII]$\lambda$3727 emission
($W_{0}([OII])$ $\ge$ 15 \AA), because we wish to compare the gas
properties to those of the underlying stellar population.

Applying these three selection criteria to the 2 CFRS fields yields a sample of
59 galaxies, of which we have chosen a small representative subsample
of 14 LCGs for a first investigation.  They represent $\sim$ 23\% of
the $M_{AB}$(B) $\le$ -20 galaxy population in the redshift range
defined above. Their global properties, including B and K luminosities
and compactness and concentration parameters, are given in Table 1.

Their apparent magnitudes range from I = 20.5 mag to 22 mag, and they
possess colors spanning the entire morphological sequence from Irr to
Sbc galaxies.  They are very similar in their properties to the
brightest $\sim$ 25\% of the blue compact galaxies selected by Guzman
et al. (1997) in the Hubble Deep Field at 0.1$<$z$<$1.3. However most
of the brightest compact galaxies of Guzman et al. have higher
redshifts than the LCGs considered here, which complicates a direct
comparison.

\section{Observations and data reduction}
\setcounter{footnote}{0}
Spectrophotometric observations of the 14 LCGs were obtained during
one night (1999 August 10-11) with the European Southern Observatory
8m VLT/UT1 using the FORS1/R600 and I600 spectrograph at a resolution
of 5\AA~ and covering the wavelength range from 5500 to 9200\AA. This
ensures the coverage from [OII]$\lambda$3727 to [OIII]$\lambda$5007 in
the rest-frame spectrum for a large fraction of the
targets. Additional exposures have been done through service observing
mode, during July and August 2000.  Each galaxy was observed through a
1" slit (Figure 2) in 5 to 7 exposures of $\sim$ 40 mn each, using the
"shift and add" technique to ensure proper removal of cosmic rays and
of possible CCD defaults.  Spectra were extracted and
wavelength-calibrated using the IRAF\footnote{IRAF is distributed by
National Optical Astronomical Observatories, which is operated by the
Association of Universities for Research in Astronomy, Inc., under
cooperative agreement with the National Science Foundation.} package.
Flux calibration was done using 15 min exposures of three different
photometric standard stars.  To ensure the reliability of the data,
all spectrum extractions as well as lines measurements using the SPLOT
program were performed independently by two of us (F.H. and N.G.).  

To disentangle the continuum properties from those of the emission lines,
we have adopted the following specific procedure for all spectra. We
applied a two-step smoothing (first by 7 pixels and then by 15 pixels)
to the continuum, while excluding from the smoothing the well
identified and mostly unresolved emission lines to keep their original
spectral resolution.  This smoothing procedure has the advantage of
not affecting the shape of the continuum and the narrow emission
lines, while revealing the more prominent absorption features. For the
latter the nominal spectral resolution is $\sim$ 10 -- 11 \AA,
comparable to that of the stellar cluster spectra used as templates in
our spectral synthesis work to be described later (section 4.2).
Figure 3 presents the 14 resulting rest-frame spectra, where the
location of strong sky emission lines (at 5577, 6300 and 6364\AA) as
well as sky absorption lines (at 6874, 7590 and 7630\AA) have been
indicated. Excluding these night sky lines, the smoothed continuum of
all spectra shows a S/N per pixel element always larger than 5. For
each spectrum, the locations of the most prominent lines are marked
and their identifications are given in the last panel.\\
\\

The $H\beta$, $H\gamma$ and $H\delta$ Balmer emission lines were found
to be systematically contaminated by absorption lines.  To disentangle
the absorption from the emission, we have synthesized the stellar
continuum by a combination of star spectra. Stellar energy
distributions were taken from the library of Jacoby et al. (1984). The
stars have solar metallicity which is appropriate for our objects as
shown later.  We used a combination of stars with spectral type
varying from B to K, and optimized the fit to the continuum for the
important spectral absorption lines, such as the high order Balmer
lines (H10 to H8), the CaII K line and the G band (Figure 4, top
panel). We impose the additional constraint that the resulting
synthesized stellar continuum not be redder than the object continuum,
since the spectra of LCGs have simultaneously strong absorption
features characteristic of old stellar populations and blue U-V colors
characteristic of young ones.  The synthesized stellar continuum is
then subtracted from the observed spectrum to yield a pure emission
spectrum uncontaminated by absorption (Figure 4, bottom panel) which
we can use to measure the Balmer decrement and derive extinction.  For
12 LCG spectra with both H$\beta$ and H$\gamma$ lines, the residual
emission lines have been measured using the SPLOT package, and the
measurement errors estimated by trying several combinations of the
stellar templates. In most cases, the Balmer emission lines are seen
to be either blue or red-shifted with respect to the center of the
absorption lines, which minimize their contamination by underlying
stellar absorption, and hence the measurement error for the gas
extinction.

Emission line widths were estimated after a careful
examination of the actual spectral resolution because seeing
conditions were varying from 0".65 to 1".0 FWHM, and the object sizes
were comparable to the slit size.  The instrumental resolution is
around 5 \AA~ (FWHM), as measured from absorption lines of the
calibration stars and unresolved sky lines.  Each line measurement was
checked against the closest sky emission line and velocity widths were
derived assuming a gaussian profile.  All spectra but one include the
[OIII]$\lambda$4959 and 5007 emission lines.  Following Guzman et
al. (1997), velocity widths derived from this set of lines are
averages, weighted by the quality of individual measurements. They are
given in Table 2.

\section{Stellar populations}
\subsection{Color properties}
Ten of the 14 LCGs possess K' band photometry (Hammer et al. 1997)
along with B, V and I photometry. At the reference redshift of z =
0.5182, the central wavelength of the observed I band is exactly that
of the rest frame V band, while the observed V band matches well the
rest frame U band centered at 3622 \AA.  At this redshift the
observed K' band represents a filter centered around 1.416
$\mu$m. Observed colors have been derived within an aperture of 3"
diameter.  Calculations of rest-frame colors has been done by
systematic fits of the observed V, I and K' photometric points, using
Bruzual and Charlot (2000) models with exponentially decreasing SFR
characterized by e-folding times $\tau$ of 0, 0.1 Gyr, 0.5 Gyr, 1.2
Gyr, 4 Gyr and infinity.  The fits have been performed without
extinction correction, and reproduce well the colors of all objects
except for CFRS 03.1349, CFRS 03.1540 and CFRS 22.1064.  The latter
objects have observed (I - K') colors which are too red even when
compared with those of a 15 Gyr old stellar population.  Because the
14 selected LCGs have redshifts close to the reference one, the
model-dependent corrections applied to the rest-frame colors (U - V)
and (V - 1.4 $\mu$m) are always negligible or smaller than the
measurements errors (i.e. $<$ 0.05 mag).  Thus, our derived rest-frame
colors are very accurate, with only a second-order dependency on the
model used to derive them.\\
\\

The resulting color distribution in Figure 5 illustrates the following
point: our sample which was constructed using a selection criterion on
the emission lines includes galaxies with (U - V) colors ranging from
those of Irr to those of Sbc, just as the sample of Guzman et
al. (1997).  On the other hand, our LCGs display a large dispersion of
the (V - 1.4 $\mu$m) color, with 7 of them showing colors much too red
compared to the predictions of the models.  Model uncertainties of the
(V - K) color -- which is very similar to the (V - 1.4$\mu$m) color --
cannot account for this effect.  These very red colors can be
explained if the presently observed burst is superimposed on a red old
stellar population, if there is significant extinction, if the stellar
population has over-solar abundances, or if there is a combination of
the above. For the reddest objects (CFRS 03.1242, CFRS 03.1349, CFRS
03.1540 and CFRS 22.1064), a combination of these effects is likely at
work. Interestingly, these four extremely red LCGs all possess a
bright companion within 4" or 36 kpc (Figure 2). These companions are
not affecting the K' band aperture photometry, except maybe for
03.1242. The CFRS 0300+00 field
was also observed by ISO to a limit of 350 $\mu$Jy (4$\sigma$) at
15$\mu$m and by the VLA to a limit of 90$\mu$Jy (5$\sigma$) at 21 cm.
Two very red LCGs, CFRS03.1349, 03.1540 in our sample have been
detected at both mid-IR and radio wavelengths.\\
\\

The dispersion in the (V - 1.4 $\mu$m) color justifies {\it a
posteriori} our adopted name for these objects, Luminous Compact
Galaxies (and not blue LCGs), since only their (U-V) colors are
blue. The lower luminosity compact galaxies of Guzman et al. are
significantly bluer in (V - 1.4$\mu$m) as compared to the LCGs in our
sample (Guzman, private communication).
\\

\subsection{Absorption lines}
All the LCG spectra show strong and broad absorption lines.  Most
prominent are the Balmer lines, the Ca I 4227\AA~ and Ca II H and K
lines, the CN band at 4182\AA, the CH band at 4301\AA, and the Mg I
line at 5173\AA~ (Figure 3). A special mention should also be made for
the important iron absorption line system in LCG spectra, including Fe
I 4037\AA~, 4264\AA~, 4384\AA~ and 4435\AA, and sometimes Fe I
4004\AA~, 4595\AA~ and 4645\AA~. FeI lines are even present below 3900
\AA~. This suggests that a large fraction of stars in LCGs have solar
abundances or greater, and that SN Ia have already released iron-peak
elements, giving an age of about 1 Gyr or more for the bulk of the
stars (see e.g. McWilliams 1997). In some LCG spectra the CN band
affects the location and shape of the H8 and H9 lines (see for example
the spectrum of 03.1540), which also suggests the presence of high
metal abundance stars.  

Table 3 gives the equivalent widths of the
most important absorption lines. The continuum has been synthesized
using a grid of stellar cluster spectra with different ages and
metallicities, and following the procedure outlined by Bica \& Alloin
(1986a and b, herafter BAa and BAb).  To test our computer routines,
we have first performed a continuum fit of the stellar cluster
spectra, and check our results against those of BAa, before applying
the procedure on LCG spectra.  Our measurement errors are slightly
lower than those of BAb 's Table 5 (from 1 to 2 \AA). We found a
correlation between the equivalent widths of the metal absorption
lines and the $(V - 1.4 \mu$m$)_{AB}$ color index (Figure 6).  This is
to be expected if the latter is mainly sensitive to age and
metallicity effects. A few objects depart from the correlation. These
are the reddest LCGs, CFRS03.1349 and 22.1064, for which we suspect
extinction to modify the $(V - 1.4 \mu$m$)_{AB}$ color index. Figure 6
also shows that the properties of most LCGs are consistent with their
stellar populations being composed of a 5 Gyr old population with
solar abundance, mixed with a young ($\le$0.1 Gyr) population, the
latter being responsible for the dilution of the absorption lines.

The Bica \& Alloin population synthesis method allows us to perform a
two-parameter analysis which takes into account both age and
metallicity effects.  Because galaxy interactions are important in
LCGs, we prefer not to adopt any assumption concerning the abundances
of the young population relative to those of the old population, as
done by Bica (1988) in his analysis of E/S0 and spiral nuclei. On the
other hand, to simplify the analysis, we have made the plausible
assumption that, in a small volume with intense star formation, the
gas is chemically homogeneous at least during the last $10^{8}$
years. Ten absorption lines have been used to perform the fit. They
include H 10 (3798 \AA), H8 (3889 \AA), Ca II K (3933 \AA), FeI
(4037 \AA), CN (4182 \AA), CaI (4227 \AA), FeI (4264 \AA), the G
band (4301 \AA), FeI (4384 \AA) and MgI (5175 \AA).  For most LCGs,
this results in 4 to 7 lines available in the spectral range, and not
contaminated by sky emission or absorption. Five star cluster
populations with ages 5 $\times$ $10^{9}$, $10^{9}$, 5 $\times$
$10^{8}$, $10^{7}$ and $10^{6}$ yr have been used as templates. We
have chosen not to use globular cluster spectra (age = 15 $\times$
$10^{9}$) because at z$\sim$ 0.5, LCGs emitted their light not more
than 6-7 Gyr ago. The metal abundance of the older population was
allowed to be one of the following values: $log(Z/Z_{\odot})$ = - 1.5,
- 1, - 0.5, 0 and 0.6. The same metal abundance has been assumed for
the $10^{9}$ and 5 $\times$ $10^{8}$ yr old populations
($log(Z/Z_{\odot})$= -1, -0.5, 0 and 0.6), and for the $10^{8}$ and
$10^{7}$ yr old populations ($log (Z/Z_{\odot})$ = -0.5, 0 and
0.6). Our goal here is not to fully disentangle the metallicity from
age effects for each LCG (our spectra do not have the required S/N to
accomplish that), but rather to have a first rough estimate of its
stellar population properties. In many cases, LCGs show large
equivalent widths of FeI 4384\AA~ and MgI 5175\AA~ and relatively
small equivalent widths of CaII K lines. The latter have been fitted
thanks to the line dilution in the blue by the younger population
which has a HII spectrum.  The fit has been done as follows: we add
first a new star cluster population to the mixture in incremental
steps of 5\% in the light contribution. If the objects are not fitted,
we reduced the step to 2\%. In some cases, the individual iron line
equivalent widths were not well reproduced by the models.  This is not
too surprising given the complexity of the Fe line system: even for
well-known stars, Fe/H is not known to better than a factor of 2 or 3
(Wheeler, Sneden \& Truran 1989). For the 6 LCGs with non-satisfactory
fits of individual Fe lines, we have fitted only the sum of the three
FeI rather than each of them individually.

Table 4 presents the result of the fits. $\chi$$^{2}$ values are
calculated as averages of the sum of the squares of the differences
between model and LCG spectrum divided by the squares of the
measurement errors. The latter are taken to be the quadratic sum of
the BAb errors and our errors.  We also restrict the parameter space
to be explored by imposing the plausible constraint that the (3728\AA~
- 5175\AA) color (approximately the (U-V) color) of the model has to
be bluer than that of the LCG. The contribution to the V light of each
population in the synthesis model is given in Table 4, in bold
characters for the major contributor. Most LCGs are dominated by
populations having ages ranging from 1 to 5 Gyr. We have tested this
result by attempting to fit the LCG absorption lines by limiting the
age range to values lower than or equal to 5$\times$$10^{8}$ yr : for
all LCGs but CFRS 22.0637 and CFRS 22.0919, the $\chi$$^{2}$ value of
the fit is well in excess of 1.  

Except for CFRS 22.0637 and CFRS
22.0919, the stellar populations in LCGs invariably include an
important contribution ($\ge$ 35\%) stars older than 1 Gyr and with
metal abundance $log(Z/Z_{\odot})$ $\ge$ -0.5. For 9 of them, we find
an important contribution of $>$ 1Gyr stars with $log(Z/Z_{\odot})$
$\ge$ 0. We have tested the robustness of the latter result by running
models where the metal abundance has been constrained to have
$log(Z/Z_{\odot})$ values lower than or equal than -0.5. We found that
for 7 LCGs, CFRS 03.0523, CFRS 03.0570, CFRS 03.0595, CFRS 03.1242,
CFRS 22.0344, CFRS 22.0429 and CFRS 22.0619, there is no possibility
to fit their absorption line spectra without an important contribution
of solar or oversolar stellar population. Special mention should be
made of CFRS 03.1242 for which we could not find a satisfactory
fit. The reason is that it shows very strong absorption lines (Table
3), stronger than the most metal-rich star clusters in our
library. Four LCGs, CFRS 03.1540, CFRS 22.0637, CFRS 22.0919 and CFRS
22.1064, apparently require low-abundance populations to fit their
absorption line spectra. However our modelling was unsuccessful to
provide physical solutions for CFRS 03.1540 and CFRS 22.1064, for
which no (or in very small amounts) young population responsible of
their emission lines have been identified. Any young component added
would dilute the absorption lines, and would lead to an old stellar
component with a solar metallicity.

While our spectral synthesis work allows us to put constraints on the
ages and metallicities of the stellar populations in LCGs, the reader
should not take the values given in Table 4 as being the actual
stellar population distributions in LCGs.  There are several caveats
which limit the results of our adopted spectral synthesis procedure.
First, there is more than one solution with $\chi$$^{2}$ values lower
than 1, and Table 4 only displays the solution with the minimum
value. These appear to be true minima as they do not change when the
incremental step for adding the light contribution of a stellar
population is changed.  Second, the fit of a galaxy population using
only a few stellar cluster templates can smooth over some important
events in the star formation history of a galaxy in which star
formation is sustained over long periods of time. We believe this
effect to be especially important in LCGs within interacting systems.
The complex nature of the star formation history and the resulting
young stellar population could then explain its absence in CFRS03.1540
and CFRS 22.1064. It can also be the reason why our modeling of CFRS
03.1242 proves to be unsuccessful.  Third, and more importantly, our
ability in disentangling age from metal abundance effects is limited
by the relatively small number of absorption lines.  For example, some
LCG spectra do not cover the blue region and do not include the
important lines below 4000\AA. This would bias the synthesis method
against finding young stellar populations which would dilute the UV
absorption lines. This effect can be seen in Table 4 for CFRS 03.1540
and CFRS 22.1064 for which the best-fit model found no young
population, although their spectra show emission lines.
What can be firmly said about the metal abundance and age of LCGs ?
First, all LCGs, but CFRS 22.0637 and CFRS 22.0919, are dominated by 1
to 5 Gyr old population, characterized in most cases by solar to
over-solar abundances.  To test the robustness of this result, we have
attempted to fit the 12 LCGs with an old stellar population with the
added constraint that the contribution of that old population does not
exceed 25\%. We found that half of the LCGs spectra could not be
fitted, their $\chi$$^{2}$ being invariably greater than 1.  As for
the remaining LCGs, they could be only fitted by assuming a young
population with the highest metal abundance available in our stellar
library ($log(Z/Z_{\odot})$ = 0.6). This is not realistic when
compared to the gas abundance values discussed in the next section.
Second, the best fits to the LCGs spectra always give a mix of stellar
populations with metal abundance covering a wide range from undersolar
(generally characterizing the youngest population) to solar or
oversolar values (characterizing the oldest population).  This is
illustrated in Figure 6, which shows that the most important lines
within a LCG cannot all be fitted with a single metallicity. As noted
above, some models of LCG spectra show no young population, in evident
contrast with their emission line spectra.  Adding a young stellar
component would inevitably enhance the metal abundance of the oldest
stellar component, making our conclusion that most LCGs are dominated
by old and solar metallicity stellar population even stronger.
 
\section{Gas properties: SFR and oxygen abundance}
For the 12 spectra with both $H\beta$ and $H\gamma$ detected, we have
been able to derive $A_{V}$ extinction values, after correcting the
Balmer emission lines for stellar absorption using the results of our
spectral synthesis work (Section 3). As shown in Table 5 and Figure 7
(left panel), the derived extinctions can have values much larger than
those currently found in present-day irregular galaxies which have an
average $A_{V}$ of $\sim$ 0.57 (Gallagher et al. 1989).  These large
extinctions imply that the SFR derived from the 2800\AA~ luminosity of
LCGs is greatly underestimated.  We have calculated for each LCG the
dereddened $H\beta$ line flux and derived the corresponding SFR using
a Salpeter IMF (Kennicutt 1998) and assuming the standard ratio
between $H\alpha$ and $H\beta$, Case B and a gas temperature of 10000
K. Figure 7 (right panel) compares the SFRs so calculated to those
derived using the 2800\AA\ continuum, assuming the same IMF (see
Kennicutt 1998).  Even allowing for errors due mainly to uncertainties
in the extinction estimates, it is clear that the UV luminosity
underestimates the actual SFR in LCGs by factors ranging from 2 to 30,
with an average of 14 and a median of 11.  The largest underestimates
are for CFRS 03.0523, CFRS 03.1540 and CFRS 03.1349, which are both
IR-luminous galaxies. From their IR and radio energy distributions,
Flores \& Hammer (2000, in preparation) derive SFRs respectively of
87, 135 and 94 $M_{\odot} yr^{-1}$.  These values are in agreement
with those derived from the extinction corrected Balmer lines,
allowing for the error bars which have large upper values. As another
consistency check, we have applied the extinction factor derived from
the $H\beta/H\gamma$ ratio to correct the SFR values at 2800\AA~ (see
Table 5).  Extinction-corrected $SFR_{2800}$ values are always
consistent with extinction-corrected $SFR_{H\alpha}$ values within the
error bars.  

The O/H abundance ratio for the gas can be estimated from
the quantity $R_{23}$ = $( [OII]3727 + [OIII]4959,5007 )/H\beta$
(Edmunds and Pagel 1984), although the relationship is degenerate and
there are two O/H values for a given $R_{23}$. The calculated $R_{23}$
for the 12 LCGs with the necessary lines range from 0.5 to 1 (Table
6), in good agreement with the values obtained for similar objects at
lower redshifts by Kobulnicky and Zaritsky (1999).  Flux measurements
of the faint [OIII]$\lambda$4363 line allow to lift the degeneracy for
two objects (CFRS22.0619 and CFRS22.0919). For those, we derive gas
temperatures of 12500 and 13000 K, close to the average upper value
for the rest of the sample (Table 6). Using Figure 8 of Kobulnicky et
al (1999), it was possible to set for each line ratio $R_{23}$, an
upper and a lower value of O/H, with the help of the ionisation ratio
$[OIII]4959,5007 /[OII]3727$.  We found that all LCGs have gas
abundances $Z/Z_{\odot}$ in the range from 0.25 to 0.4, with the
exception of CFRS 22.0919, to be discussed later, which shows a very
low O/H.  Adopting a statistical correction of +0.1 dex on the O/H
abundance as suggested by Kobulnicky et al. (1999) for physical
parameters derived from global galaxy spectra, our data are finally
consistent with a median metal abundance $Z/Z_{\odot}$ = 0.45, similar
to the value derived by Kobulnicky and Zaritsky (1999) for their
objects.  Thus, gas in LCGs displays oxygen abundances intermediate
between those of local spirals ($Z/Z_{\odot}$ $\sim$ 1) and those of
irregular dwarf galaxies ($Z/Z_{\odot}$ $\sim$ 0.3).

\section{Discussion}
We have obtained VLT spectra of 14 LCGs selected by their compactness
and luminosity in the rest-frame blue.  The blue luminosities of LCGs
range from 0.35 to 1.4 $L^{\star}$ with an average of 0.70
$L^{\star}$.  They are also bright in the near-IR, with a rest-frame
K-band luminosity ranging from 0.1 to 2.8 $L^{\star}$ and an average
of 0.86 $L^{\star}$, assuming $M_{K}^{\star}$(AB)=-22.55 for $H_{0}=50$ (Glazebrook
et al, 1995).  Guzman (1999) has noticed that LCGs are also compact at
near-IR wavelengths.  Here we assume that the K-light samples the
stellar mass (Charlot 1998), with a mass-to-IR light ratio very close
to unity. This follows Charlot (1996) who has investigated this ratio
for a solar abundance population with various ages, appropriate for
LCGs. A Salpeter Initial Mass Function (IMF) has been adopted, as it
allows models to reproduce well the evolution of the global IR and B
luminosity densities from z = 0 to z =1, to which the LCGs contribute
importantly. On the other hand, Lilly et al. (1996) found that the use
of a Scalo IMF in modeling the z = 0 -- 1 galaxy population produces
too many long-lived low-mass stars.  Also photometric properties of
local disks suggest a Salpeter rather than a Scalo IMF (Kennicutt et
al. 1994). In any case, the adopted 0.22 dex uncertainty based on
Charlot (1996)'s study of the conversion of K luminosity to mass
likely accounts for the major uncertainties related to the IMF. 

LCGs have stellar masses ranging from $10^{10}$ to 2.5 $\times 10^{11}$
$M_{\odot}$, with an average of 7 $\times 10^{10}$ $M_{\odot}$ (see
Table 1). K-band luminosities derived from the 1.4$\mu$m luminosities
are robust as the (1.4$\mu$m-K) color shows small variations during
the period from $10^{8}$ years (it has the value -0.32 mag) to
$10^{10}$ years (its value is -0.15 mag) after the beginning of star
formation.  Uncertainties in these derivations are small compared to
the large uncertainty involved in the conversion of the K-light into
mass ($\sim$ 0.22 dex in Figure 8). From the way they have been
selected and their properties, the LCGs studied here are very similar
to the 25\% most luminous blue compact galaxies identified by Guzman
et al (1997) in the Hubble Deep Field (HDF).  They are the galaxies
responsible for the decrease in the UV luminosity density since the
epoch corresponding to z=1 (Lilly et al. 1998).  

With the noticeable
exception of CFRS 22.0919 and CFRS 22.0637, LCGs all display strong
metallic absorption lines and have (U-V) colors similar to those of
late-type galaxies, in the range Sbc to Irr. These properties can be
reproduced by a combination of an old star population with age $\ge$
$10^{9}$ yr with a younger population with age $\le$ 5 $\times$
$10^{8}$ yr (see Section 4.2).  It is important to note that the best
spectral synthesis fits are generally obtained with an older stellar
population with abundances larger than or equal to solar values, while
the younger component has solar or sub-solar abundances.  The latter
component diminishes the observed strength of the absorption lines due
to the older stellar population, as well as making the $(U-V)_{AB}$
color bluer than that of a Sbc galaxy.  Analysis of the emission lines
indicates that the ionized gas in LCGs also displays oxygen abundances
in the range from 0.3 to 0.5 the solar value, i.e. the metal abundance
of the gas is similar to that of the younger stellar population.

Given these observations, what can we say about the nature of LCGs?
The presence of a moderately old stellar population with solar
abundances or larger, stellar masses near that of the Milky Way, large
SFRs similar to or only slightly below those of luminous IR galaxies,
all argue against a scenario in which LCGs are progenitors of today's
spheroidal or irregular dwarf galaxies as discussed by Koo et
al. (1995) and Guzman et al. (1997).  That hypothesis is based mainly
on one key observation: that of the small velocity widths of the
emission lines of LCGs (Table 2). These authors argue that the
velocity widths are representative of virialized motions of the gas
and since the velocity widths are similar to those of dwarf galaxies,
they are dwarf progenitors.  However an examination of the LCG spectra
( Figure 3), shows that the Balmer emission lines are often
red-shifted relatively to the Balmer absorption, suggesting that more
complex mechanisms may be responsible of the narrowness of their
widths. We also note that, in some cases, emission lines have
asymetric profiles. In the case of CFRS 03.0523, they show two peaks
separated by $\sim$ 470 km s$^{-1}$, which are likely to be associated
with each of the two merging components (Gruel et al.  2001, in
preparation). A special comment should be made about CFRS 22.0637 and
CFRS 22.0919 which, as we have noted before, appear to be different
from other LCGs and resemble more HII galaxies. In particular, CFRS
22.0919 has a young low-metallicity stellar population as well as a
low gas oxygen abundance, it shows a moderate gas extinction, and a
large velocity difference (800 km s$^{-1}$) between the emission and
absorption lines.  Gas outflow can be responsible for the blue-shifted
Balmer absorption lines, so that CFRS 22.0919 may be just the type of
dwarf progenitor object discussed by Koo et al. (1995) and Guzman et
al. (1997). 

But objects such as CFRS 22.0919 constitute more the
exception than the rule in our LCG sample.  The stellar mass as well
as the light concentration of other LCGs is more reminiscent of those
of bulges of today's Sab to Sbc spirals or even ellipticals. Could
LCGs be the progenitors of bulges of present-day massive galaxies?
The bulk of their stars displays a large range in metal abundance, as
in the Milky Way's bulge. Taking into account their large extinctions,
LCGs are found to be still forming stars at a very vigorous rate, with
a median SFR equal to 40 $M_{\odot} yr^{-1}$.  For 8 LCGs with the
necessary data, the median characteristic formation time scale for the
bulk of the stars derived from the emission lines is $\sim$ 1.8 Gyr
(Figure 8).  This value is consistent with that obtained by the
absorption line analysis, but is much larger than the characteristic
time of $\sim$ 0.1 Gyr for the monolithic scenario of bulge formation
(Elmegreen 1999). Thus, the latter scenario does not appear to be the
appropriate one for explaining the observed properties of the majority
of LCGs.  

Which mechanism can then explain the transformation of LCGs
to present-day spirals?  An important hint can be obtained from HST
images.  These show that an important fraction, if not all, of the
LCGs possess a low-surface-brightness component which surrounds their
core (Figure 2, see also Guzman, 1999 or Hammer, 1999). Moreover, a
substantial fraction of them (6/14) are seen to have bright companions
within a radius of 40 kpc. The LCGs with companions show the largest
SFRs in the sample, and they are also the reddest in
(V-1.4$\mu$m). The three galaxies in the sample which have been
detected by ISO (Flores et al. 2000, in preparation) are among
them. It is thus plausible to think that gas tidally pulled out from
the companion and falling into a LCG can feed the star formation
within it. This would naturally explain the metal abundance difference
between the gas and the oldest stars.  A good example of a LCG where
such a mechanism may be at work is CFRS 03.1540: its
low-surface-brightness companion acts as a reservoir of low abundance
gas to sustain the star formation in it during several $10^{8}$ yr.
CFRS 03.1540 also shows a well-developped disk -- it was classified as
a disk galaxy by Lilly et al. (1998) -- and star formation is probably
occurring in both bulge and disk. The tadpole galaxies described by
Brinchmann et al.  (1998) which would have been selected as LCGs by
our criteria, can be interpreted in this context as undergoing the
first stage of the formation of a disk around the bulge. Thus, by
observing LCGs, we are witnessing an important phase of galaxy
formation: the building of the bulge and the beginning of the disk
formation of $L^{\star}$ spirals, where galaxy interactions play a
major role.  

The scenario developped here has to account for the
narrow emission line widths reported by Phillips et al. (1997) and
also seen in several of our spectra.  In normal galaxies, the bulge is
dynamically relaxed, and the gas emission likely samples most of the
velocity field, i.e. the emission line widths should be indicative of
the whole galaxy potential.  Weedman (1983) reported however unusually
narrow emission line widths in starburst nuclei. Lehnert and Heckman
(1996) studying a sample of IR-selected starburst galaxies, have shown
that the starburst occurs preferentially in the inner region, so that
the gas does not sample fully the solid body part of the rotation
curve. They found several objects with emission line velocity widths
of less than 50 km s$^{-1}$ in systems with rotational velocities
ranging from 100 to 200 km s$^{-1}$.  Since they have SFRs and hence
IR luminosities, equal or slightly below those of ISO galaxies (Flores
et al. 1999), we conclude that our LCGs are very similar to some
Lehnert \& Heckman galaxies, and that the narrow emission line widths
are not indicative of the entire potential of the LCGs and lead to a
systematic underestimate of their true masses.  

The present study thus
suggests that LCGs are progenitors of today's massive (non-dwarf)
galaxies, which have formed the bulk of their stars at z $<$ 1.  We
note that:\\ 
1) the gas supplied to LCGs from interacting companions
is enough to sustain their star formation during several $10^{8}$ yr,
until completion of the bulge stellar content as well as to feed a
gradual formation of the disk.\\ 
2) the characteristic time for the
formation of the bulk of LCG stars is relatively small, averaging to
only $\sim$ 1.8 Gyr, i.e. most of the stars formed at z $<$ 1.\\
Because LCGs represent $\sim$ 20\% of the field galaxies, the fraction
of stars in massive galaxies formed at relatively recent epochs could
have reached 20\%, and even more if the star formation in interacting
systems of large (non-compact) galaxies detected by ISO (Flores et
al. 1999; Hammer, 1999) is taken into account.  This estimate is much
higher than the one obtained by Brinchmann \& Ellis (2000), who find
the fraction to be negligible.  The main reason for the discrepancy
comes from the way the SFR is derived.  Brinchmann and Ellis (2000)
calculate the SFR from the [OII]3727 flux, while we use the
extinction-corrected $H\alpha$ emission, assuming the standard ratio
between $H\alpha$ and $H\beta$, Case B and T = 10000 K. Our derived
SFRs can be a factor 10 or more greater than those derived by
Brinchmann \& Ellis (2000). This can be seen by comparing their Figure
3 to our Figure 8. Using the same sample as that of Brinchmann \&
Ellis (2000), Hammer and Flores (1998) have shown that SFRs derived
from either [OII]3727 or 2800\AA~ luminosities do not correlate with
those obtained from $H\alpha$ luminosities, mainly because of
extinction effects.

Given their large extinction and
$H\alpha$-derived SFRs, it is likely that the IR luminosities of LCGs
are similar to or only slightly below those of the luminous IR
galaxies detected by ISO in the same redshift range (Flores et
al. 1999). Because they represent 20\% of the field galaxies in the
redshift range from 0.4 to 0.8, the contribution of the LCGs to the
global SFR density needs to be reevaluated. Such a reevaluation will
probably lead to an increase in the fraction of the SFR density
related to photons reprocessed by dust, i.e. there should be an
evolution in the shape of the IR luminosity function, due to the
addition of new populations -- such as LCGs -- around $L_{IR} $=$
10^{11}L_{\odot}$. This subject cannot be addressed properly with our
present small sample.  We shall discuss it in a future paper with a
much enlarged sample.

\section{Conclusions}
Spectroscopic observations of a small sample of 14 Luminous Compact
Galaxies (LCG) with the VLT have provided new and important
information on a crucial stage in the formation of galaxies, that of
bulges in massive (non-dwarf) spiral galaxies. The LCG spectra are
characterized by the following properties:

1) they show strong metallic absorption lines, including those
from $\alpha$-elements and iron.

2) they show large Balmer decrement ratios (as measured by $H\beta/H\gamma$), 
implying large extinctions (median $A_V$ of $\sim$ 1.5 mag).

3) the SFRs derived from the extinction-corrected Balmer lines are more
than ten times higher than the corresponding ones derived from the UV 
luminosities.

LCGs with redshifts between 0.4 and 0.8 are undergoing intense bursts
of star formation, with SFRs averaging to $\sim$ 40 $M_{\odot}
yr^{-1}$.  Because of their large extinctions, they are expected to be
luminous in the IR, at a luminosity level just below the IR luminous
galaxies detected by ISO in the same redshift range (Flores et
al. 1999).  Indeed 3 of the 14 LCGs discussed in this paper are
detected by ISO, implying higher extinction for them.

We believe that LCGs are progenitors of present-day bulges of massive
spiral galaxies because:

1) they have stellar masses similar to those of today's bulges in 
$\ge$$L^{\star}$ spirals, concentrated in comparable volumes.

2) their stars show a wide range of metal abundance, from sub-solar
to over-solar values, as in the bulge of the Milky Way.

3) the majority of LCGs show low-surface-brightness components around
their high-surface-brightness cores (the bulges). Most are in
interacting systems, and the tidally pulled gas from the companions is
likely to sustain star formation in the LCGs at a high rate for
several hundreds Myr, resulting in the completion of the bulge
formation and the beginning of the disk formation.  The characteristic
time for forming the bulk of the stars in LCGs is rather small, only
$\sim$ 1.6 Gyr on average. It is comparable to the age of the oldest
stars in LCGs derived from the analysis of their absorption lines.
This means that a significant fraction (several tens of percent) of
present day's massive galaxies could have been formed at relatively
recent epochs, at z $\le$ 1.  This is consistent with the cosmic star
formation history which predicts that an important fraction of the
stellar mass was formed in the z = 0 -- 1 redshift range (Hammer
1999).  The importance role played by galaxy interactions in the
formation of bulges of spiral galaxies is also consistent with the
hierarchical theory of galaxy formation. On the other hand, the
scenario outlined here may appear to be in contradiction with the
results of Lilly et al. (1998) who found that the number density of
large spirals at z $\sim$ 1 is comparable to that today. However the
Lilly et al. results are based on only $\sim$ 40 morphologically
selected large spirals in the redshift range from 0.2 to 1, and their
large one $\sigma$ error bar can easily accomodate a density evolution
of spiral disks of $\sim$ 20\% as found here.  Moreover, some of the
large spirals in Lilly et al. (1998)' sample are in interacting
systems and are strong infrared emittors, such as CFRS 03.1540
discussed here, or other examples discussed by Flores et al. (1999).
Presumably, these large interacting disks will merge leading to a
density evolution.

  If the 14 LCGs described here are representative of the whole LCG
  population at z $> $0.4, then their contribution to the global star
  formation density should be revised upwards by factors as large as 7
  -- 10. That contribution could then be as high or higher than that
  of the luminous IR galaxies detected by ISO (Flores et al. 1999).
  The latter are more massive systems and formed stars more rapidly
  than LCGs, but they constitute a much sparser population of galaxies,
  comprising only a few percent of the total population. Most LCGs
  at z $\sim$ 0.5 would have bolometric IR luminosities not far below
  the ISO sensitivity limit, and we predict that they will be easily
  detected in large numbers by SIRTF.

 Finally we reiterate that our interpretation of the nature of LCGs is
 quite different from that proposed by Koo et al. (1995), Guzman et
 al. (1997) and Guzman (1999) for the majority of compact galaxies
 found in the Hubble Deep Field and flanking fields. This is probably
 because most of the Guzman et al galaxies have intrinsically lower
 luminosities than LCGs.  We believe LCGs to be 10 to 100 times more
 massive than dwarf galaxies, to be forming stars at large rates, so
 they cannot be the progenitors of local spheroidal or irregular
 systems through fading. Corresponding to the brightest 25\% of the
 compact galaxies selected by Guzman et al (1997), the LCGs selected
 in CFRS fields are mostly evolved starbursts probably similar to
 those classified as such in the HDF. Due to their large derived SFRs,
 LCGs are the main contributors to the large SFR density observed at
 z=0.5 to 1.

 In future papers (Gruel et al. in preparation) we will investigate
 the velocity fields and dynamics of LCGs, using data of superior
 quality and based on a larger sample. We will also be studying the
 possible revision of the star formation history and of the IR
 luminosity density caused by the LCG population.


\acknowledgments
We are grateful to C. Balkowski, F. Combes, R. Guzman, D. Kunth and
C. Vanderriest for useful discussions and advices. We are especially
indebted to Pascale Jablonka who introduced us to the methodology of
fitting absorption line spectra by star cluster synthesis. We thank
the ESO Allocation Time Committee for the one night attributed to this
program at UT1. F.H. is very grateful to Claire Moutou and Thomas
Szeifert for their help at Paranal and their patience.  We thank the
data flow operations team for the fast delivery of the data.  We thank
the CNRS/CONICYT for financial support of our collaboration.
T.X.T. is grateful for the hospitality of the Departement d'Astronomie
Extragalactique et de Cosmologie at the Observatoire de Meudon and the
Institut d'Astrophysique during his sabbatical leave. He thanks the
partial financial support of the Centre National de la Recherche
Scientifique, of the Universit\'e of Paris VII and of a
Sesquicentennial Fellowship from the University of Virginia.


\begin{table*}[htbp]
\tablenum{1}
\caption{Luminosity and compactness}
\begin{center}

\begin{tabular}{ccrrrrrrrrrc} \hline 
\\
CFRS    &      z   &   M$_B$  &  $\frac{L_B}{L_\star}$ & Err    &   M$_K$    &  $\frac{L_K}{L_\star}$ & 
Err     &  $L_{K}$$^1$ & $\delta$I$_{CFHT}^2$ &  $\delta$I$_{HST}^2$   &    log(r$_{1/2})^3$      \\ 
\\ \hline   
03.0442 &  0.47807 &  -20.30  &  0.436  & 0.064  & -21.06  &  0.254 &  0.084  &  2.2 & 0.68 &    --  &     --   \\   
03.0523 &  0.65355 &  -21.20  &  1.000  & 0.147  & -22.24  &  0.752 &  0.194  &  6.5 & 0.43 &   0.46 &    0.58  \\   
03.0570 &  0.64682 &  -20.38  &  0.470  & 0.117  &   ---   &   ---  &   ---   &  --- & 0.30 &    --  &     --   \\
03.0595 &  0.60442 &  -20.78  &  0.680  & 0.119  & -22.09  &  0.655 &  0.157  &  5.6 & 0.31 &   0.54 &    0.65  \\   
03.0645 &  0.52737 &  -20.77  &  0.673  & 0.049  & -22.20  &  0.724 &  0.113  &  6.3 & 0.62 &   0.64 &    0.69  \\   
03.1242 &  0.76786 &  -21.21  &  1.009  & 0.223  & -22.99  &  1.500 &  0.317  & 12.9 & 0.73 &    --  &     --   \\   
03.1349 &  0.61640 &  -21.43  &  1.236  & 0.114  & -23.67  &  2.805 &  0.206  & 24.2 & 0.73 &   0.35 &    0.52  \\   
03.1540 &  0.68931 &  -21.52  &  1.343  & 0.124  & -22.83  &  1.294 &  0.274  & 11.2 & 9999 &   0.56 &    0.67  \\   
22.0344 &  0.51680 &  -20.30  &  0.437  & 0.048  &   ---   &   ---  &   ---   &  --- & 0.62 &    --  &     --   \\   
22.0429 &  0.62433 &  -20.29  &  0.433  & 0.083  & -22.13  &  0.679 &  0.144  &  5.9 & 0.57 &    --  &     --   \\   
22.0619 &  0.46706 &  -20.08  &  0.356  & 0.082  &   ---   &   ---  &   ---   &  --- & 0.75 &    --  &     --   \\   
22.0637 &  0.54188 &  -21.26  &  1.057  & 0.107  & -21.83  &  0.515 &  0.099  &  4.4 & 0.61 &    --  &     --   \\   
22.0919 &  0.47144 &  -20.22  &  0.406  & 0.034  &   ---   &   ---  &   ---   &  --- & 9999 &   0.37 &    0.42  \\   
22.1064 &  0.53685 &  -20.06  &  0.350  & 0.074  & -22.43  &  0.895 &  0.214  &  7.7 & 0.21 &    --  &     --   \\ \hline  
  \\ 
\end{tabular}
  \leavevmode
\end{center}
Notes:\\
$^1$ in units of $10^{10}$ $L_{\odot}$\\
$^2$ light concentration, $m_{I}(15h_{50}^{-1}kpc)$ - $m_{I}(5h_{50}^{-1}kpc)$; 9999: unsuccessful measurement\\
$^3$ $r_{1/2}$ in kpc\\
\end{table*}

\begin{table*}[htbp]
\tablenum{2}
\caption{FWHM of emission lines}
\begin{center}
\begin{tabular}{ccrcrcrcr} \hline 
\\
CFRS    & FWHM$_{OII}^1$ &  Q$^2$  & FWHM$_{H\gamma}^1$ &  Q$^2$  &  FWHM$_{OIII}^1$   &	 Q$^2$  &	FWHM$^1$ & Q$^2$ \\ 
\\ \hline
03.0442 &      --  &     9  &      50     &  3  &      50     &  1  &      50    &   1\\
03.0523 &     115  &     2  &     150     &  2  &      180    &  1  &     180    &   1\\
03.0570 &      50  &     3  &      79.5   &  3  &      130    &  1  &     130    &   1\\
03.0595 &      54  &     4  &     150     &  2  &       50    &  2  &      50    &   2\\
03.0645 &      50  &     2  &      50     &  2  &       55    &  2  &      55    &   2\\
03.1242 &     135  &     3  &      81     &  3  &       --    &  9  &     108    &   2\\
03.1349 &     252  &     2  &      60     &  2  &      250    &  2  &     250    &   1\\
03.1540 &      50  &     3  &      50     &  3  &       60    &  2  &      60    &   1\\
22.0344 &      --  &     9  &      50     &  2  &       50    &  2  &      50    &   2\\
22.0429 &      52  &     4  &      50     &  3  &       50    &  2  &      50    &   2\\
22.0619 &      50  &     3  &      50     &  2  &       50    &  1  &      50    &   1\\
22.0637 &      89  &     4  &      50     &  2  &       50    &  1  &      50    &   1\\
22.0919 &      --  &     9  &      50     &  1  &       50    &  1  &      50    &   1\\
22.1064 &      --  &     9  &      76     &  2  &       82    &  2  &      80    &   1\\ \hline
  \\ 
\end{tabular}
  \leavevmode
\end{center}
Notes:\\
$^1$ In [km/s] \\
$^2$ Quality factor: (1) error of 10\%, (2) error of 20\%, (3) error of 50\% and (9) not measured  \\
\end{table*}

\begin{table*}[htbp]
\tablenum{3}
\caption{Absorption lines }
\begin{center}
\begin{tabular}{ccccccccccc} \hline
\\
CFRS    &       H10     &       H8      &       CaIIK   &       FeI     &       CN      &       CaI     &       FeI     &       G       &       FeI     &       MgI+MgH \\
\\ \hline
$\lambda$&       3795   &       3890    &       3933    &       4137    &       4182    &       4229    &       4264    &       4301    &       4392    &       5176 \\
\hline
03.0442 &       9999.0  &       9999.0  &       9999.0  &       1.825   &       4.121   &       2.232   &       9999.0  &       9999.0  &       5.500   &       9999.0  \\
03.0523 &       9999.0  &       5.262   &       8.413   &       0.878   &       9999.0  &       1.537   &       1.972   &       2.372   &       8.273   &       9999.0  \\
03.0570 &       11.172  &       9999.0  &       13.042  &       3.825   &       9999.0  &       6.507   &       8.937   &       6.227   &       10.147  &       9999.0  \\
03.0595 &       9999.0  &       10.948  &       9999.0  &       2.843   &       7.734   &       2.543   &       9999.0  &       9999.0  &       6.808   &       9999.0  \\
03.0645 &       8.185   &       9999.0  &       7.354   &       9999.0  &       9999.0  &       1.462   &       3.613   &       5.967   &       5.868   &       9999.0  \\
03.1242 &       6.647   &       9999.0  &       6.758   &       6.176   &       14.247  &       7.236   &       10.440  &       9999.0  &       13.837  &       9999.0  \\
03.1349 &       2.672   &       9999.0  &       9999.0  &       2.031   &       4.828   &       3.384   &       9999.0  &       3.662   &       3.640   &       9999.0  \\
03.1540 &       9999.0  &       7.031   &       5.142   &       2.660   &       2.510   &       3.945   &       3.836   &       6.328   &       7.587   &       9999.0  \\
22.0344 &       9999.0  &       9999.0  &       9999.0  &       9999.0  &       9999.0  &       3.206   &       3.202   &       7.177   &       9.421   &       8.239   \\
22.0429 &       4.091   &       9999.0  &       9999.0  &       2.591   &       5.497   &       9999.0  &       0.314   &       3.806   &       9.394   &       9999.0  \\
22.0619 &       9999.0  &       3.185   &       5.668   &       0.898   &       5.983   &       3.679   &       4.410   &       9999.0  &       6.773   &       9999.0  \\
22.0637 &       2.596   &       4.662   &       2.309   &       9999.0  &       1.107   &       0.691   &       0.880   &       0.833   &       9999.0  &       9999.0  \\
22.0919 &       9999.0  &       9999.0  &       9999.0  &       3.414   &       1.569   &       1.751   &       9999.0  &       9999.0  &       3.373   &       4.118   \\
22.1064 &       9999.0  &       9999.0  &       9999.0  &       9999.0  &       3.782   &       4.684   &       4.969   &       3.799   &       3.081   &        6.119  \\
\hline
  \\ 
\end{tabular}
  \leavevmode
\end{center}
Notes:\\
9999.0: line or continum point used to fit the line is outside the spectral range, or the
line is contaminated by the sky
\end{table*}

\begin{table*}[htbp]
\tablenum{4}
\caption{LCG stellar populations}
\begin{center}
\begin{tabular}{cccccccccc} \hline
\\
CFRS	&    ${\chi}^2$	&	N$^1$	&	age		&$5 \times 10^9$&	$10^9$	&$5 \times 10^8$&$5 \times 10^7$&	$10^7$	&	$10^6$	\\ 	
\hline
\hline
03.0442	&	0.029	&	4	&	$Z/Z_{\odot}$$^2$&	-1.0   	&	-0.5   	&	---   	&	-0.5   	&	-0.5     &	---	\\
	&		&		&	$\%$		&	15     	&    \bf{75}  	&	0     	& 	5     	&	5      &	0	\\
\hline
03.0523	&	0.250	&	5	&	$Z/Z_{\odot}$	&	---     &	0.6	&	---	&	---	&	-0.5	&	---	\\
	&		&		&	$\%$		&       0	&    \bf{50}    &	0      	&	0     	&    \bf{50} 	&	0	\\
\hline
03.0570	&	0.361	&	5	&	$Z/Z_{\odot}$	&	0.6    &	0.6	&	---	&	-0.5	&	---	&	---	\\
	&		&		&	$\%$		&	10	&    \bf{80}	&	0      	&	10     	&	0	&	0	\\
\hline
03.0595	&	0.229	&	5	&	$Z/Z_{\odot}$	&	-1.5    &	0.6    	&	---   	&	---   	&	-0.5    	&	---	\\
	&		&		&	$\%$		&	25	&    \bf{60}	&      	0     	&	0       &	15     	&	0	\\
\hline
03.0645	&	0.471	&	6	&	$Z/Z_{\odot}$	&	-1.5   	&	0.6   	&	---   	&	---	&	-0.5    	&	0.0	\\
	&		&		&	$\%$		&	35	&    \bf{50}	&      	0	&     	0	&      	10     	&	5	\\
\hline
03.1242	&	1.530	&	6	&	$Z/Z_{\odot}$	&	---   	&	0.6   	&	---   	&	---   	&	-0.5   	&	---	\\
	&		&		&	$\%$		&	0	&    \bf{95}	&	0     	& 	0     	&	5     	&	0	\\
\hline
03.1349	&	0.286	&	5	&	$Z/Z_{\odot}$	&	-1.5    &	0.6 	&	---   	&	---   	&	---    	&	0.0	\\
	&		&		&	$\%$		&     	\bf{50}	&       30	&	0     	& 	0     	&	0     	&	20	\\
\hline
03.1540	&	0.915	&	6	&	$Z/Z_{\odot}$	&	-1.5   	&	-0.5	&	---   	&	---   	&	---	&	---	\\
	&		&		&	$\%$		&     	10	&    \bf{90}	&	0     	& 	0     	&	0	&	0	\\
\hline
\\ 
\end{tabular}
  \leavevmode
\end{center}
Notes:\\
$^1$ Number of absorption lines used for the fit\\
$^2$ in logarithmic unit
\end{table*}	

\begin{table*}[htbp]
\tablenum{4 (continued)}
\caption{LCG stellar populations }
\begin{center}
\begin{tabular}{cccccccccc} \hline
CFRS	&    ${\chi}^2$	&	N$^1$	&	age		&$5 \times 10^9$&	$10^9$	&$5 \times 10^8$&$5 \times 10^7$&	$10^7$	&	$10^6$	\\ 	
\hline
\hline
22.0344	&	1.836	&	5	&	$Z/Z_{\odot}$	&	-1.5	&	0.6	&	---   	&	---   	&	-0.5	&	---	\\
	&		&		&	$\%$		&	15	&    \bf{70}	&	0	&	0	&	15	&	0	\\
\hline
22.0429	&	0.100	&	4	&	$Z/Z_{\odot}$	&	---    	&	0.6    	&	---   	&	---	&   	-0.5    &	0.0	\\
	&		&		&	$\%$		&      	0	&    \bf{55}	&      	0      	&	0     	&	40     	&	5	\\
\hline
22.0619	&	0.420	&	5	&	$Z/Z_{\odot}$	&	0.6    &	-1.0     &	---	&	---	&	---     &	0.0	\\
	&		&		&	$\%$		&     \bf{55}	&    	10	&	0	&	0	&	0	&	35	\\
\hline
22.0637	&	0.296	&	7	&	$Z/Z_{\odot}$	&	-1.5	&	-1.0	&	---	&	---	&	-0.5    &	0.0	\\
	&		&		&	$\%$		&	35	&	20      &	0      	&	0	&     \bf{55}	&    	10	\\
\hline
22.0919	&	0.167	&	4	&	$Z/Z_{\odot}$	&	---   	&	-1.0   	&	---	&   	-0.5   	&	-0.5	&	---	\\
	&		&		&	$\%$		&     	0 	&    \bf{90} 	&     	0     	&	5	&     	5	&	0	\\
\hline
22.1064	&	0.466	&	5	&	$Z/Z_{\odot}$	&	---	&	-0.5	&	---   	&	---   	&	-0.5	&	---	\\
	&		&		&	$\%$		&	0	&     \bf{95}	&      	0      	&	0      	&	5     	&	0	\\
\hline
\\ 
\end{tabular}
  \leavevmode
\end{center}
Notes:\\
$^1$ Number of absorption lines used for the fit\\
$^2$ in logarithmic unit
\end{table*}	

\begin{table*}[htbp]
\tablenum{5}
\caption{Extinction and SFR}
\begin{center}
\begin{tabular}{crrrrccl} \hline 
\\
CFRS    & $\frac{H\beta}{H\gamma}$ &  Err   &    A$_v$    &   Err   & SFR$_{2800}^1$   &  SFR$_{H\alpha}^2$   & Err    \\ 
\\ \hline   
03.0442 &  2.64  &   0.52 &    1.53  &   $^{+1.30}_{-1.53}$  	&   1.52  &   11.54	&  $^{+11.91}_{-2.70}$  \\   
03.0523 &  2.68  &   0.24 &    1.65  &   $^{+0.62}_{-0.68}$  	&   4.20  &  107.35$^3$  	&  $^{+63.05}_{-31.90}$ \\   
03.0570 &  2.18  &   0.43 &    0.15  &   $^{+1.30}_{-0.15}$    	&   1.80  &    3.27  	&  $^{+6.52}_{-1.48}$   \\
03.0595 &  2.45  &   0.48 &    1.00  &   $^{+1.30}_{-1.00}$   	&   2.00  &   18.70  	&  $^{+32.23}_{-7.30}$ \\   
03.0645 &  2.64  &   0.17 &    1.54  &   $^{+0.46}_{-0.49}$  	&   3.33  &   37.76  	&  $^{+9.34}_{-5.67}$   \\   
03.1242 &   --   &    --  &    --    &    --   			&   ---   &     --   	&     --   		\\   
03.1349 &  2.98  &   0.45 &    2.42  &   $^{+1.01}_{-1.18}$  	&   3.79  &  134.38$^3$ &  $^{+151.16}_{-48.25}$\\   
03.1540 &  3.04  &   0.36 &    2.56  &   $^{+0.82}_{-0.93}$  	&   4.39  &   88.50$^3$ &  $^{+90.79}_{-36.20}$ \\   
22.0344 &  2.76  &   0.22 &    1.86  &   $^{+0.56}_{-0.61}$  	&   1.67  &   27.57  	&  $^{+8.61}_{-4.64}$   \\   
22.0429 &   --   &    --  &     --   &    --                 	&    --   &     --   	&      --   		\\   
22.0619 &  2.29  &   0.22 &    0.51  &   $^{+0.66}_{-0.51}$  	&   1.04  &    9.62  	&  $^{+3.00}_{-1.46}$   \\   
22.0637 &  2.37  &   0.19 &    0.76  &   $^{+0.56}_{-0.61}$	&   3.69  &    5.25  	&  $^{+1.84}_{-0.99}$   \\   
22.0919 &  2.25  &   0.08 &    0.38  &   $^{+0.27}_{-0.28}$	&   2.41  &    8.42  	&  $^{+0.86}_{-0.64}$   \\   
22.1064 &  2.78  &   0.19 &    1.91  &   $^{+0.48}_{-0.51}$  	&   1.61  &   37.66  	&  $^{+10.25}_{-6.09}$  \\ \hline  
  \\ 
\end{tabular}
  \leavevmode
\end{center}
Notes:\\
$^1$ SFR in [M$\odot/yr$], without correction for extinction\\
$^2$ SFR in [M$\odot/yr$] corrected from extinction calculated from the $H\beta$/$H\gamma$ ratio\\
$^3$ also detected by ISO (Flores et al, 2000, in preparation).\\
\end{table*}

\begin{table*}[htbp]
\tablenum{6}
\caption{Gas properties }
\begin{center}
\begin{tabular}{ccrcrcrrrr} \hline
\\
CFRS   & $log(R_{23})^1$ & R$_{ion}^2$& OIII ratio$^3$  & T$_{gaz}$[K]   &  e(T)$^4$  &    $\frac{O}{H}_1^5$  
& $\frac{O}{H}_2^5$ &  $\frac{O}{H}$   & Err \\
\\ \hline
03.0442 &	1.10  	&	0.22  &	65      &  	15000  	&    3	&  8.2	&  8.4	& 8.3	&  0.1 \\
03.0523 &	0.76 	&	0.57  &	233	&   	11000  	&    3  &  8.0  &  8.5  & 8.3   &  0.2 \\
03.0570 &	0.69 	&	0.76  &	149	&	12000  	&    3	&  7.9  &  8.7  & 8.3   &  0.5 \\ 
03.0595 &	0.31 	&	0.30  & 50      &       18000 	&    3  &  7.6  &  8.9  & 8.3   &  0.6 \\     
03.0645 &	0.79 	&	0.56  & 111	&       12500 	&    3	&  8.3	&  8.55	& 8.4   &  0.15\\
03.1349 &	0.60 	&	0.59  &	174	&	11500  	&    3  &  8.0  &  8.8  & 8.4   &  0.4 \\
03.1540 &	0.57 	&	0.05  & 30      &       22000  	&    3  &  8.2  &  8.9  & 8.5   &  0.3 \\
22.0344 &	0.67 	&	0.31  &	119     &       12500  	&    3  &  8.0  &  8.8  & 8.4   &  0.4 \\
22.0619 &	0.87 	&	1.54  &	116     &       12500	&    2	&  8.3	&  8.2	& 8.2	&  0.05\\
22.0637 &	0.72 	&	0.58  & 75      &       14500  	&    3  &  8.2  &  8.6  & 8.4   &  0.2 \\
22.0919 &	0.86 	&	33.24 &	105	&       13000	&    1	&  7.8	&  8.8	& 7.8	&  0.05\\
22.1064 &	0.73 	&	0.97  & 40	&       21000   &    3	&  7.85	&  8.8	& 8.3	&  0.5 \\
\hline
  \\ 
\end{tabular}
  \leavevmode
\end{center}
Notes:\\
$^1$ See text.  \\
$^2$ Ionization  \\
$^3$ $[OIII]5007+[OIII]4959/[OIII]4363$ ratio\\
$^4$ Quality factor: 1 excellent, 2 estimated, 3 temperature limit \\
$^5$ O/H : 1 lower value and 2  higher value (cf. Kobulnicky, 1999) \\
\end{table*}	

\clearpage


\begin{figure*}
\begin{center}
\includegraphics [width=15cm] {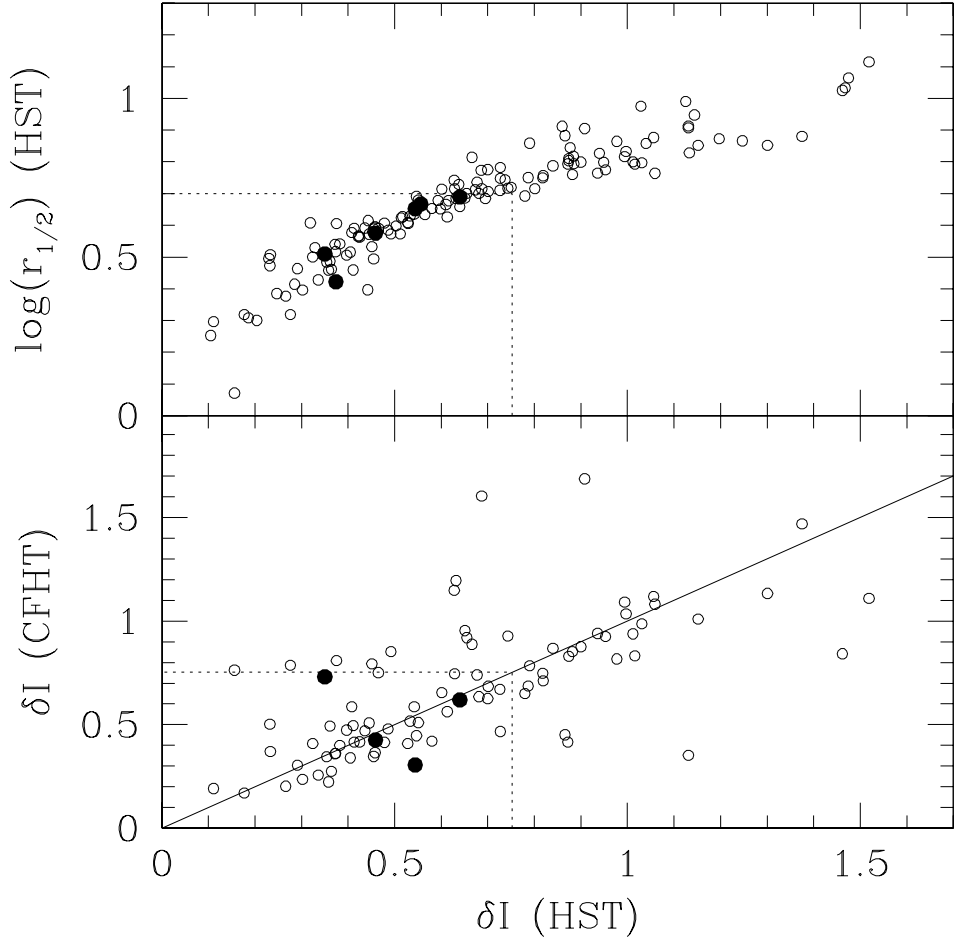} 
\end{center}
\caption{ Compactness selection criteria for Luminous
Compact Galaxies (LCG).  {\it Top}: logarithm of half-light radius
r$_{1/2}$ (in kpc) against light concentration parameter for galaxies
observed with HST/WFPC2 in the F814W filter; filled dots show the 6
LCGs (CFRS 03.523, CFRS 03.0595, CFRS 03.0645, CFRS 03.1349, CFRS
03.1540 and CFRS 22.0919) in this paper imaged with the HST. The
dotted lines delimit the compact galaxy area. CFRS 03.523 is made of
two very close components and was classified as a merger by Le F\`evre
et al (2000).  {\it Bottom}: comparison between the light
concentration parameter derived from ground-based images (FWHM = 0".7)
and that derived from HST images. CFRS 03.1540 and CFRS 22.0919 have
ground-based images which aperture magnitudes showing a decrease in
brightness with the aperture radius decreasing from 3" to 0.2". Their
concentration parameter cannot be derived, probably because of the
presence of a companion which was not properly masked.  In the absence
of HST imagery, those two galaxies would have been excluded from our
sample.}
\end{figure*}

\begin{figure*}
\begin{center}
\end{center}
\caption{ {\bf JPEG enclosed}
{\it Left}: Ground-based Canada-France-Hawaii telescope images of the
14 LCGs in the sample, with the slit superimposed.  {\it Right}: The 6 
LCGs in the sample observed by WFPC2/HST. HST image quality reveals
low surface brightness extents and closeby interactions , generally not 
detectable from the ground.}
\end{figure*}

\begin{figure*}
\begin{center}
\end{center}
\caption{ {\bf JPEG enclosed}
Rest-frame VLT spectra of the 14 LCGs in our sample. 
The continuum has been smoothed first by 7 pixels and then by 15
pixels, giving a spectral resolution of 11 \AA, except at the
location of the emission lines ([OII]3727, $H\epsilon$, $H\delta$,
$H\gamma$, $H\beta$, [OIII]4959 and 5007) for which the rest-frame
spectral resolution is 3.5 \AA.  Pairs of vertical dashed lines
delimit the regions where strong sky emission lines (OI 5577, 6300 and
6364 \AA) or absorption lines (O2 6877, 7606 and 7640 \AA) severely
affect the continuum. The main absorption lines are indicated and
their identifications are given in the last panel. The latter shows an
enlargement of the CFRS 03.1540 spectrum, as well as the adopted fit
for its continuum (solid line), used to calculate the absorption line
equivalent widths (see section 4.2).}
\end{figure*}

\begin{figure*}
\begin{center}
\includegraphics [width=15cm] {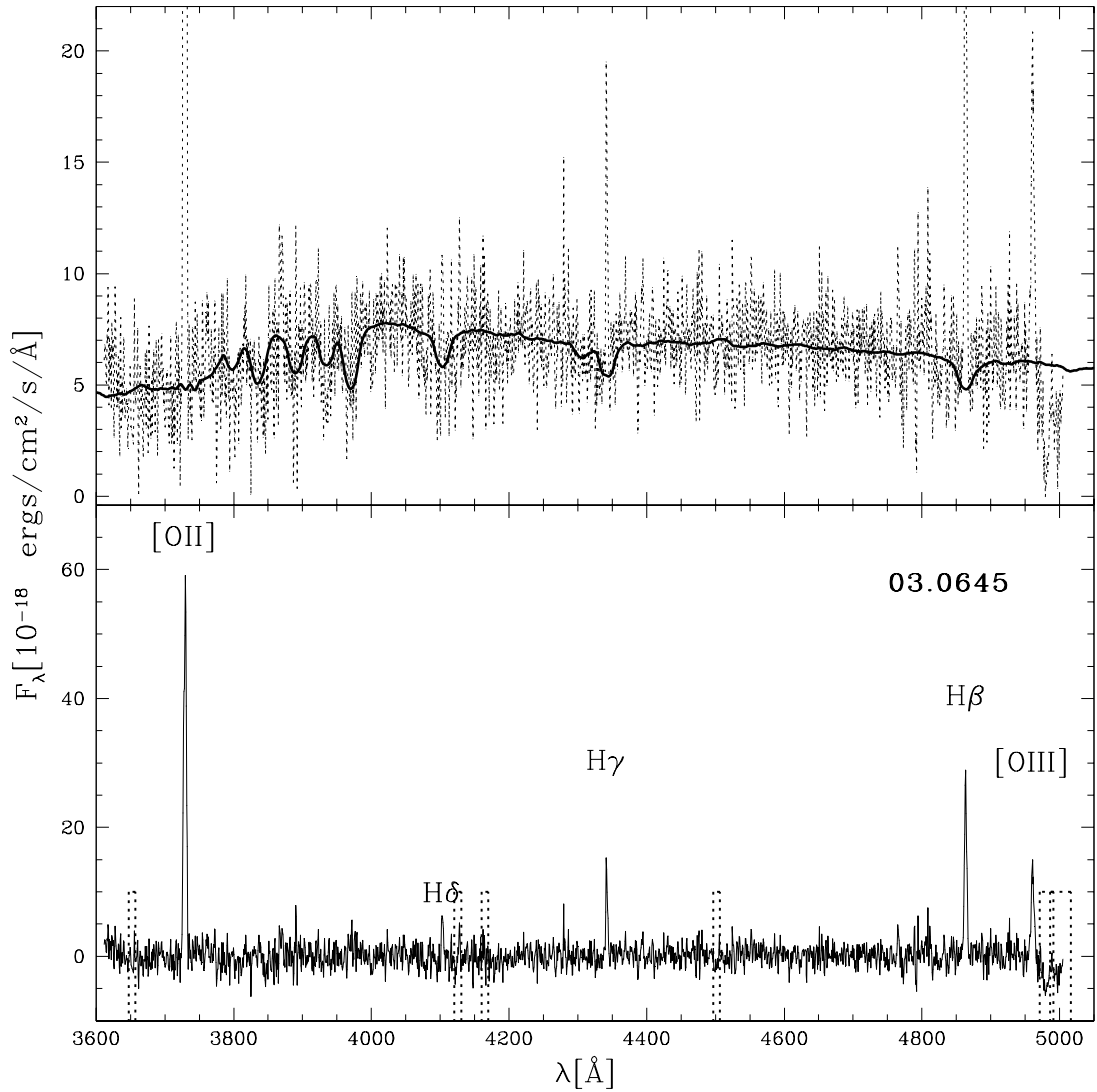}
\end{center}
\caption{{\it Top}: Rest-frame spectrum of CFRS 03.0645 
without smoothing on which a synthesized spectrum resulting from the
combination of B V (contributing 22\% of the light), A V (20\%), F V
(42\%) and G V stars (16\%) is superimposed (bold line).  {\it
Bottom}: emission line spectra of CFRS 03.645 after removal of the
stellar continuum as determined from spectral synthesis.  Pairs of
vertical dashed lines delimit the regions where strong sky emission or
absorption affect the continuum.}
\end{figure*}

\begin{figure*}
\begin{center}
\includegraphics [width=15cm] {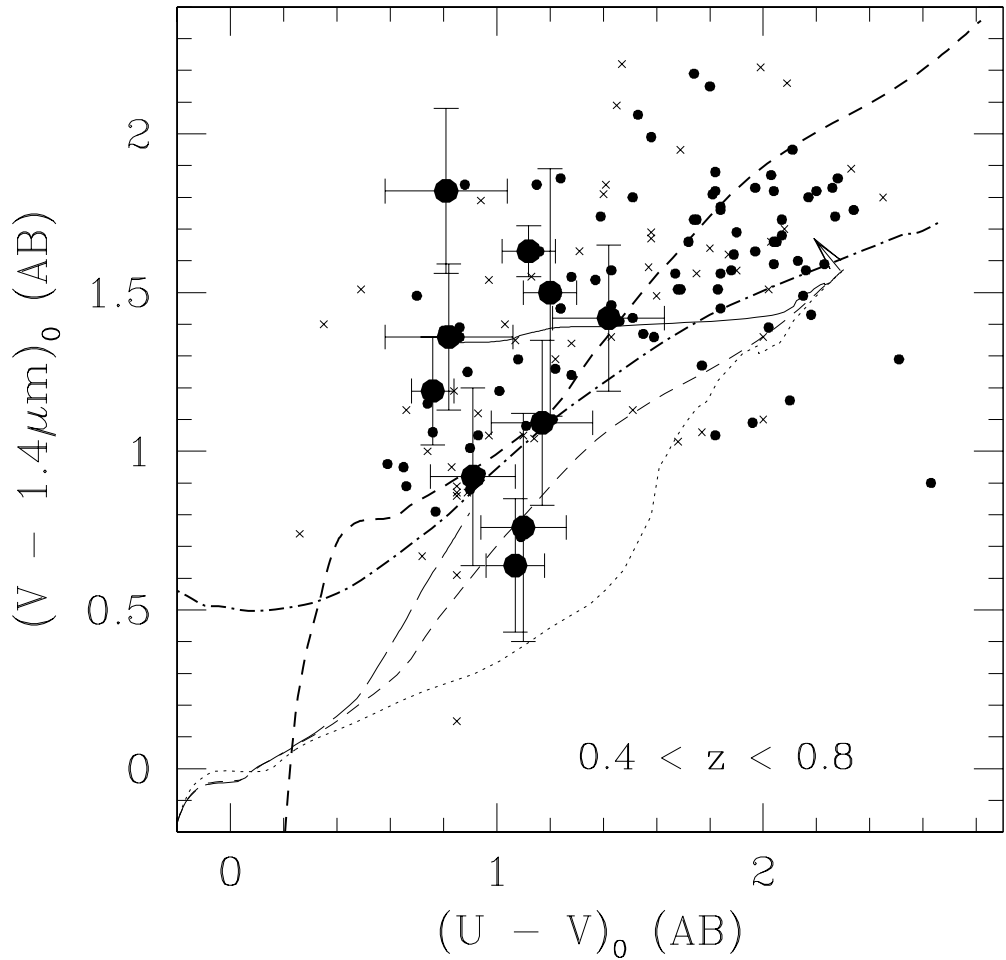}
\end{center}
\caption{Rest-frame color $(V - 1.416 \mu$m$)_{AB}$ versus 
$(U - V)_{AB}$ for CFRS luminous galaxies ($M_{B} \le$ -20).  Small
dots represent LCGs with or without emission lines, including
ellipticals. Small crosses represent galaxies which are not
compact. Large dots show the 10 galaxies in our sample of 14 LCGs
which have near-IR photometric measurements.  Error bars are mainly
due to observational errors (see text).  Lines show Bruzual and
Charlot (1999) models with solar metallicity and an exponential
declining SFR with an e-folding time-scale $\tau$.  The dotted line
show the model with $\tau$ = 0.1 Gyr, the short-dashed line the model
with $\tau$ = 1 Gyr and the long-dashed line the model with a constant
SFR. The bold short-dashed line represents the $\tau$=1 model with an
extinction $A_{V}$ = 1 applied. The bold dot-short dashed line
represents the $\tau$ = 1 Gyr model, assuming a metallicity
$Z/Z_{\odot}$ = 3.2.  A composite model, where 80\% of the light is
contributed by an old stellar population formed 10 Gyr ago in an
instantaneous burst, and the remaining 20\% of the light by a younger
stellar population with $\tau$ = 1 Gyr, is shown as a solid line. It
can be seen that, because of their large $(V-1.4\mu$m$)_{AB}$, several
LCGs in our sample require either extinction, population mixing,
over-solar abundances or a combination of the above.  A small arrow on
the upper-right side of the Bruzual and Charlot (1999) models
indicates the uncertainties in the modeling of the old stellar
population, as discussed by Charlot et al. (1996).}
\end{figure*}

\begin{figure*}
\begin{center}
\includegraphics [width=15cm] {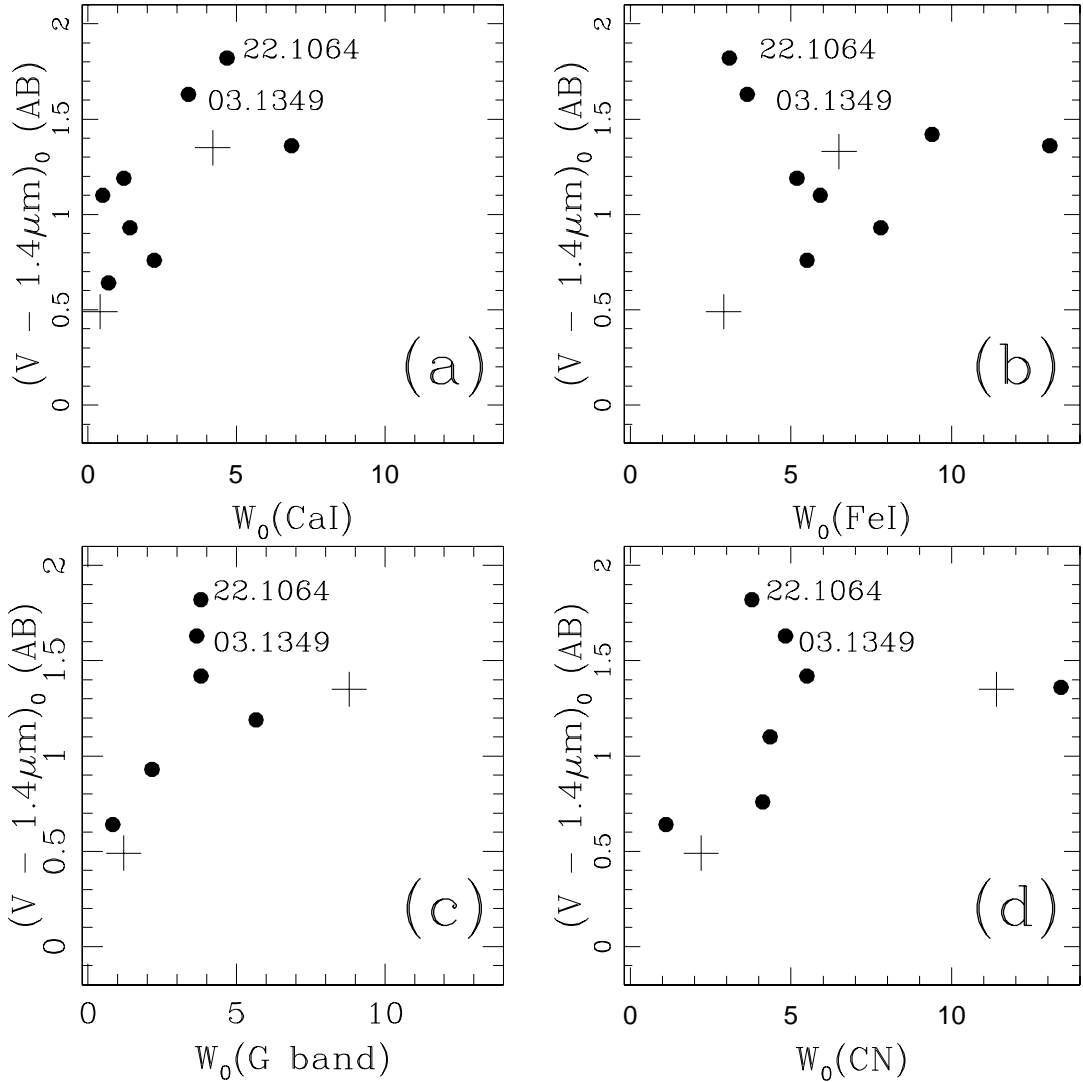}
\end{center}
\caption{$(V-1.4\mu$m$)_{AB}$ color index against 
equivalent widths of a) CaI 4227\AA, b) FeI 4384\AA, c) CH (G band
at 4301\AA) and d) CN band at 4182\AA.  Red $(V-1.4\mu$m$)_{AB}$
colors are related to strong metallic absorption lines, i.e.  either
to large abundances or to old stellar populations.  At least two
galaxies, CFRS 03.1349 and CFRS 22.1064, deviate from the
relationship, probably because their color indices are significantly
affected by extinction.  In each panel the two crosses represent the
properties of stellar clusters with solar metallicity, with ages of 5
Gyr (upper right) and 0.1 Gyr (bottom left) respectively. Comparison
of the locations of the filled dots relative to the crosses in the
four panels illustrates the fact that most LCG properties can be
fitted by a mix of a solar abundant old stellar population with a
young population.}
\end{figure*}

\begin{figure*}
\begin{center}
\includegraphics [width=15cm] {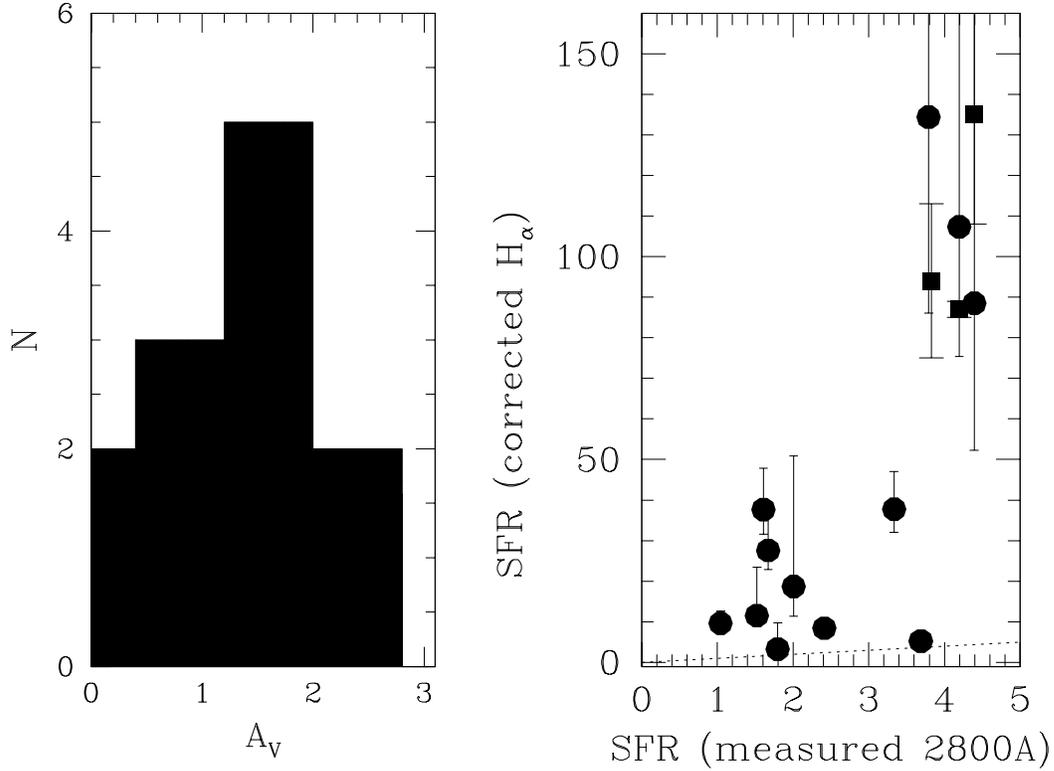}
\end{center}
\caption{{\it Left}: 
Histogram of the extinction coefficient $A_{V}$ as derived from the
$H\beta/H\gamma$ ratio. The signal-to-noise ratio of the $H\gamma$
line ranges from 10 to 30, which leads to a 1 $\sigma$ error of 0.5 to
1 mag on $A_{V}$ (see Table 5).  {\it Right}: The SFRs derived from
dereddened Balmer emission lines compared to those derived from
2800\AA~ luminosities not corrected for extinction.  Vertical error
bars represent 1$\sigma$ uncertainties, mostly due to the extinction
corrections.  For three LCGs, CFRS 03.0523, CFRS 03.1349 and CFRS
03.1540, the global SFR has been also calculated from the fit of their
spectral energy distribution from UV to radio wavelengths, including
ISO mid-IR measurements (see text).  These three estimates are
represented by squares, which are consistent with estimates from
extinction corrected Balmer lines (full dots located at the same
$SFR_{2800}$ value).}
\end{figure*}

\begin{figure*}
\begin{center}
\includegraphics [width=15cm] {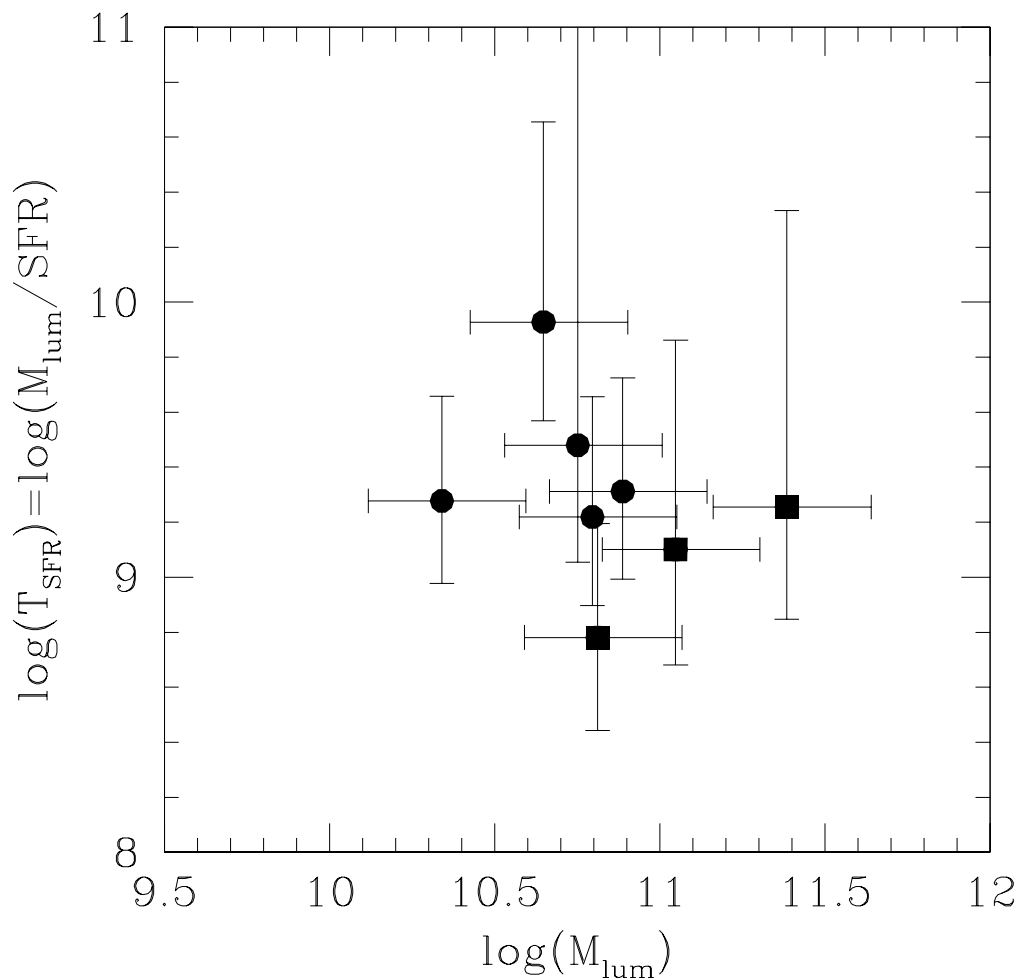}
\end{center}
\caption{Logarithm of the ratio between the luminous mass and 
the global SFR against the logarithm of the luminous mass. The former
ratio is the required time T$_{SFR}$ to form the bulk of the LCG
stars, assuming that star formation is sustained at the observed
value.  Luminous masses have been derived from the K-band luminosity
while the SFRs have been derived either from the dereddened Balmer
emission line fluxes or by fitting the global spectral energy
distribution (squares).  A mass-to-light ratio $M_{K}/L_{K}$ = 1 has
been assumed to derive the luminous mass. Vertical error bars
represent the full range of values that this ratio can take, from 0.6
to 1.8, assuming a burst older than 0.1 Gyr (Charlot 1999).
Horizontal error bars show the 1$\sigma$ error on the SFR
estimate.}
\end{figure*}


\end{document}